\documentstyle[amstex,aps,prb,epsfig,floats]{revtex}

\begin{document}

\wideabs{ 
\title{First-principles study of nucleation, growth, and interface structure of Fe/GaAs}
\author{Steven C. Erwin,$^{1,2}$ Sung-Hoon Lee,$^{2,}$\cite{permanentaddress} and Matthias Scheffler$^{2}$}
\address{$^1$Center for Computational Materials Science, Naval Research Laboratory, Washington, D.C. 20375}
\address{$^2$Fritz-Haber-Institut der Max-Planck-Gesellschaft, Faradayweg 4--6, D-14195 Berlin-Dahlem, Germany}
\date{\today}
\maketitle

\begin{abstract}
We use density-functional theory to describe the initial stages of Fe
film growth on GaAs(001), focusing on the interplay between chemistry
and magnetism at the interface. Four features appear to be generic:
(1) At submonolayer coverages, a strong chemical interaction between
Fe and substrate atoms leads to substitutional adsorption and
intermixing. (2) For films of several monolayers and more, atomically
abrupt interfaces are energetically favored. (3) For Fe films over a
range of thicknesses, both Ga- and As-adlayers dramatically reduce the
formation energies of the films, suggesting a surfactant-like
action. (4) During the first few monolayers of growth, Ga or As atoms
are likely to be liberated from the interface and diffuse to the Fe
film surface. Magnetism plays an important auxiliary role for these
processes, even in the dilute limit of atomic adsorption. Most of the
films exhibit ferromagnetic order even at half-monolayer coverage,
while certain adlayer-capped films show a slight preference for
antiferromagnetic order.
\end{abstract}

\pacs{PACS numbers: 75.70.-i,82.65.+r,75.50.Bb,72.25.Mk }
}

\section{Introduction}

Magnetic materials are widely anticipated to be integrated into
semiconductor-based microelectronics during the next decade or two.
\cite{prinz98}  A major component of this effort has focused on using
ferromagnetic thin films as a source for creating spin-polarized
electrical current in a semiconductor substrate, a process referred to
as ``spin injection.'' Three criteria have been identified as
important for technologically useful spin injection: (1)
Substantial spin polarization of the injected current; (2) Electrons,
rather than holes, serving as the spin-polarized carriers; (3) Curie
temperatures for the source of order room temperature or higher.

Although several materials appear to meet one or two of these
criteria, none has yet met all three.  Several diluted magnetic
semiconductors based on Be- and Mn-doping of
ZnSe,\cite{fiederling99,jonker00} and Mn-doping of GaAs,\cite{ohno99}
have recently demonstrated spin-injection efficiencies of greater than
50\%, but only near liquid helium temperatures. Other magnetic
semiconductors, including CdCrSe, can be doped both $p$- and $n$-type
and have Curie temperatures of order 100--200 K,\cite{haas70} but
their performance as spin injection sources has yet to be examined.

In addition to the magnetic semiconductors, much current research 
continues to focus on one of the earliest studied candidate materials, Fe.
\cite{prinz81}  Besides offering the possibility of room-temperature
injection of electron spins, Fe has the potential advantage of forming
a nearly lattice-matched epitaxial film on an important semiconductor
substrate, GaAs.  However, despite this early promise and after
considerable research investment, measured spin-injection efficiencies
for Fe/GaAs remain frustratingly low, typically no larger than 1\%.
\cite{hammar99,xu99}
The reason for this low efficiency is not definitively known, and much
controversy surrounds its origin.  A crucial issue, not yet settled,
is whether the measured efficiencies reflect an intrinsic upper limit,
or are simply due to technical limitations that may be overcome or
circumvented.  For example, much early effort focused on the nature of
interface layers inferred to be ``magnetically dead'', in the sense
that they do not to contribute to the total magnetic moment of the Fe
film.\cite{krebs87,filipe97} The appearance of dead layers is
consistent with the formation (thermodynamically favorable in the
bulk) of nonmagnetic FeAs complexes at an interface with the As-rich
GaAs(001)-(2$\times$4) surface. This reasoning led to the development
of two strategies for suppressing As diffusion into the Fe film:
passivation of the As-rich surface by a surfactant such as
sulfur;\cite{anderson95} and growing on the {\it Ga-rich}
GaAs(001)-(4$\times$6) surface (or, equivalently, As decapping prior
to growth).\cite{zoelfl97} Both approaches lead to films with
magnetization onset in the range 4--8 monolayers and with essentially
the full moment per Fe atom in all
layers.\cite{zoelfl97,kneedler97a,xu98}

Notwithstanding the successful elimination of dead layers, measured
spin-injection efficiencies have, until recently, remained at or below
the 1\% level. Recently, Schmidt {\it et al.\@} have argued that a
more fundamental limitation exists for spin injection from a
ferromagnetic metal into a semiconductor.\cite{schmidt00} They have
shown that in the purely diffusive regime (where spins are scattered
much less frequently than electrons) the spin-injection efficiency
from a ferromagnet (fm) into a semiconductor (sc) is proportional to
the ratio of their conductivities, $\sigma_{\rm sc}/\sigma_{\rm fm}$.
For ferromagnetic metals this ratio is of order 10$^{-4}$, and for
typical device geometries suggests maximum injection efficiencies of
1\% or less.

There remain several possibilities for circumventing this limitation
on Fe sources.  The first is to operate in the ballistic regime, where
the contact resistance due to elastic backscattering at the interface
(Sharvin resistance) will generally reduce the metal-semiconductor
conductivity mismatch.\cite{schmidt00} Tang {\it et al.\@} have used
concepts from mesoscopic transport to model injection into a
two-dimensional electron gas, and find clear evidence for ballistic
spin transport that otherwise vanishes in the diffusive
limit.\cite{tang00}

A second possibility, relevant to the diffusive regime, is the use of
tunnel contacts at the Fe/GaAs interface, which are expected to
substantially reduce the conductivity mismatch and thereby increase
the injection efficiency.\cite{rashba00} Related research avenues
concern the role of intrinsic Schottky barriers in controlling the
spin-dependent tunneling through an interface.\cite{hirohata00}
Indeed, Zhu {\it et al.} have recently demonstrated injection from Fe
into GaAs with an efficiency of about 2\% at room temperature; they
suggest that the Schottky barrier formed between Fe and GaAs leads to
a tunnel contact, thus circumventing the problems of conductance
mismatch.\cite{zhu01} Very recently, Hanbicki {\it et al.\@} have used a
Schottky contact to an AlGaAs overlayer to inject spin into GaAs with
efficiencies of 30\% at low temperature and 9\% at room
temperature.\cite{hanbicki02}

Since the prospects and limitations for spin injection from
Fe into GaAs remain uncertain, we believe that further progress may
benefit from a first-principles theoretical description of the
interface. Our focus will be on the atomic structure of the interface
and its resultant magnetic character, especially during the first few
monolayers of growth.  Theoretical studies of buried interfaces are
notoriously difficult, for several reasons. First, experimental probes
can provide only indirect information about physical and electronic
structure, and hence are of limited utility for guiding
theories. Moreover, real interfaces may---even after careful
annealing---have atomic geometries very different from the ground
state, hence the interface structure may depend on the precise growth
history. In principle, one approach to this dilemma would be
first-principles finite-temperature molecular dynamics simulation of
film growth, which would properly account for the roles of deposition
rate, surface diffusion, and incorporation into the substrate. In
practice, molecular dynamics using density functional theory can at
best simulate processes for $\sim$100 picoseconds---many orders of
magnitude short of the experimentally relevant time scales, which may
be milliseconds or longer.  Kratzer and Scheffler have recently
addressed this problem using a ``first-principles kinetic Monte
Carlo'' method.\cite{kratzer01} However, the complexity of applying
this method to the growth of Fe on GaAs---with three atomic species
and many different possible processes---makes such an approach not yet
feasible.

\begin{figure}[tb]
\centerline{\epsfig{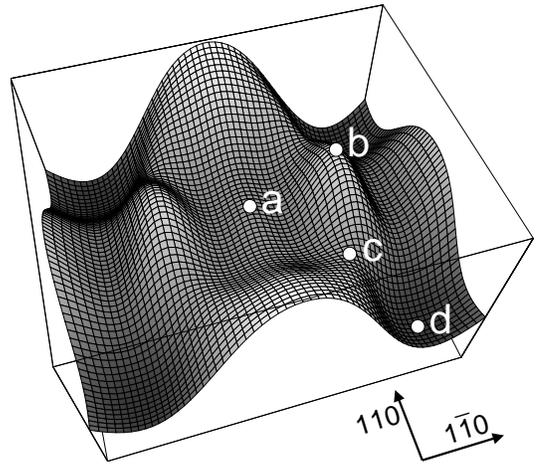}}
\caption{\label{pes.fe1.lda}
Potential energy surface for Fe on GaAs(001), calculated
using GGA without spin polarization. The labeled points correspond to the
configurations in Fig.~\protect\ref{xmol3d.reactioncoords}.}
\end{figure}

In this paper we approach the problem from two different limits. In
Section \ref{Initial_stages_of_growth} we consider the initial stages
of interface formation, beginning with adsorption of isolated Fe
adatoms on a bare GaAs substrate. We focus here on the potential-energy
surface governing surface diffusion, and show that Fe-As chemistry may 
play a decisive role in the submonolayer regime. In Section
\ref{Ground_state_interface_structure} we consider a different limit,
namely the thermodynamic ground-state interface structure of Fe films
several monolayers thick. Here we concentrate on the magnetic
character of the interface as a function of film thickness, and
propose a mechanism that accounts for the experimentally observed
delayed onset of ferromagnetic order for films of a few
monolayers.\cite{zoelfl97,xu98,kneedler97b} In Section
\ref{Summary} we discuss ways to reconcile our results from 
these two limiting regimes, and we propose future directions for
further study.

\section{Initial stages of growth}\label{Initial_stages_of_growth}

\begin{figure}[tb]
\centerline{\epsfig{figure=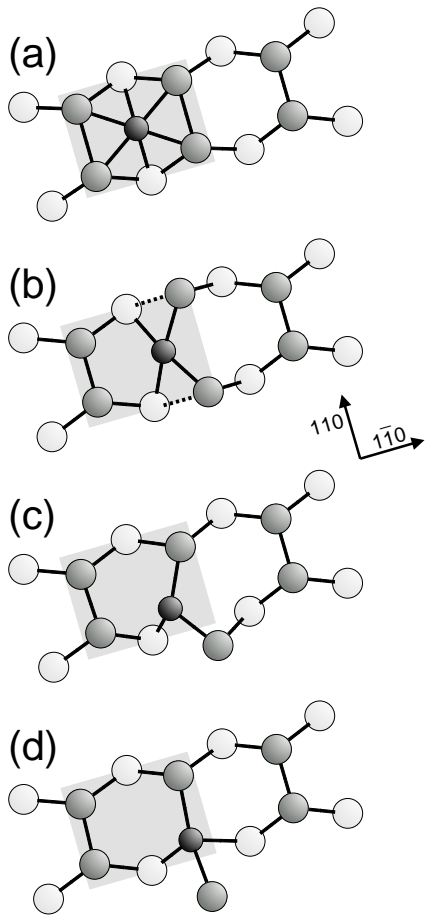,width=6cm}}
\caption{\label{xmol3d.reactioncoords}
Relaxed configurations for Fe adsorbed at the four locations
marked in Fig.~\protect\ref{pes.fe1.lda}. The gray and light gray
spheres represent Ga and As atoms, and the smaller dark gray spheres
represent Fe atoms. Highly strained bonds (in the range 15--20\% longer
than bulk) are shown as dotted lines.
The shaded area marks the region for which the
potential-energy surface is plotted in
Fig.~\protect\ref{pes.fe1.lda}. }
\end{figure}

Our goal in this section is to identify and quantify those structural,
magnetic, and chemical features that may be {\it generic} to the
growth of Fe on GaAs. We do not attempt a definitive treatment of
growth on a particular GaAs reconstruction or at a particular
temperature.  Instead, we focus on the following four questions: (1)
Does the initial adsorption of atomic Fe prefer metallic bonding at
highly coordinated sites, or does covalent bonding to Ga or As
prevail? (2) Does adsorbed Fe nucleate the formation of compact
islands, or do adsorbate-substrate interactions dominate the growth?
(3) Is surface diffusion of Fe likely to be significant at growth
temperatures?  (3) Are the Fe magnetic moments in the submonolayer
regime governed by strong Hund coupling (atomic moments) or by
itinerancy effects (bulk moments)?

We begin by studying Fe adsorption on a fictitious, but chemically
reasonable, surface of GaAs(001):  a ($2\times 1$) reconstruction
consisting of
bulk GaAs terminated by a dimerized Ga layer.
This fictitious surface is analogous to the dimerized
$\beta(2\times 4)$ and $\beta 2(2\times 4)$ As-rich surfaces, but is
quite different from the more complex $\zeta(4\times 2)$ surface
believed to be the basis for the $c(8\times 2)$ reconstruction 
observed under Ga-rich conditions.\cite{lee00}

\begin{figure}[tb]
\centerline{\epsfig{figure=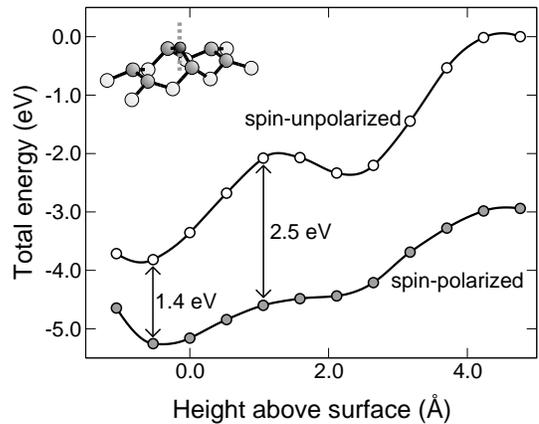,width=7.0cm}}
\caption{\label{et-vs-z}
Relative total energy for Fe constrained to approach the Ga-dimer
bridge site from directly above, as shown.  Energies are with respect
to the spin-unpolarized energy for Fe far above the GaAs surface. The
vertical distance between the two curves is the magnetic energy
gain. Without spin polarization there is a local barrier of $\sim$0.4
eV for breaking the Ga dimer; when spin polarization is included this
barrier is removed.}
\end{figure}

We use density-functional theory together with the generalized
gradient approximation (GGA) for the exchange-correlation
functional.\cite{perdew92} For the calculations in this section we
used Troullier-Martins pseudopotentials and a plane-wave basis with a
kinetic-energy cutoff of 50 Ry, as implemented in the {\sc fhi96md}
code.\cite{bockstedte97} The sampling of the surface Brillouin zone
was equivalent to using 64 $k$-points for a ($1\times 1$) surface unit
cell.  

To construct the potential-energy surface (PES) for adsorption of
atomic Fe on the clean surface, we computed the total energy, $E_t$,
as a function of the adsorbate position $(X,Y)$ within the surface
unit cell. To minimize interactions of the adsorbate with its periodic
images we used a ($2\times 2$) supercell. For each adsorbate position
$(X,Y)$, we fully relaxed the $Z$ coordinate of the adsorbate and
the positions of all Ga and As atoms in the top two
layers of the surface. This procedure was repeated for adsorbate
positions sufficient to sample the PES with a resolution of 
about 0.25 \AA. For many adsorbate positions, geometries with subsurface
adsorption or with atomic positions exchanged were considered as well;
in each case, the lowest-energy configuration was used to define
$E_t(X,Y)$.  

\begin{figure}[tb]
\centerline{\epsfig{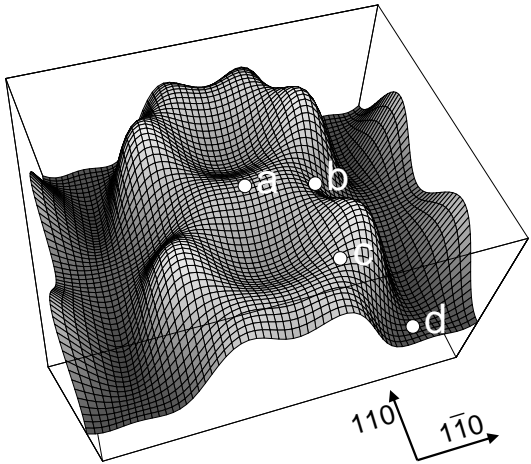}}
\caption{\label{pes.fe1.lsda}
Potential energy surface for Fe on GaAs(001), calculated
with spin polarization. The labeled points are the same as in
Fig.~\protect\ref{pes.fe1.lda}, and correspond very closely to the
configurations in Fig.~\protect\ref{xmol3d.reactioncoords}. }
\end{figure}

In order to illuminate the detailed role---if any---of magnetism in
the adsorption energetics, we begin by first calculating the PES
without allowing for spin-polarization.  We then recalculate the PES
while allowing for spin-polarization, and examine the differences that
arise.  If the adsorbate moment were completely localized and did not
interact with the substrate, these two energy surfaces would be
identical except for an overall shift of the energies.  Thus, the
differences that we find (described below) directly reflect
site-dependent magnetic interactions between the Fe moment and the
GaAs substrate.

Fig.~\ref{pes.fe1.lda} shows the PES calculated without spin
polarization, for a portion of the surface unit cell shown in
Fig.~\ref{xmol3d.reactioncoords}.  We restrict the plot to the
vicinity of the dimer rows because the energy landscape between dimer
rows is quite flat and considerably higher in energy.  Along the dimer
rows the PES is periodic with repeat length $a/\sqrt{2}$;
Fig.~\ref{pes.fe1.lda} shows slightly more than one full period along
this direction.

Each point on this PES corresponds to a different relaxed geometry,
determined solely by the in-plane adsorbate position $(X,Y)$. We focus
on the four points marked in Fig.~\ref{pes.fe1.lda}, which correspond
to the four geometries shown in
Fig.~\ref{xmol3d.reactioncoords}. Point (a), in the center of the PES,
corresponds to adsorption at the high-symmetry pedestal site, midway
between two dimers along a dimer row. Although this site is highly
coordinated, leading to four Fe-Ga and two Fe-As bonds of nearly equal
length (2.5 \AA), it is energetically unfavorable.  Indeed, this site
is not even locally metastable: energy is gained by moving the
adsorbate in any direction away from point (a).

The global minimum of the PES is a point (d) near the corner of the
region shown in Fig.~\ref{pes.fe1.lda}.  This is 0.8 eV below point
(a) and corresponds to the configuration of
Fig.~\ref{xmol3d.reactioncoords}(d), in which the Fe adsorbate has
partially ``kicked out'' one Ga atom from a surface dimer and taken
its place. The large energy gain from this process suggests that the
formation of such Fe-Ga heterodimers may act as a strong local trap
for Fe, strongly suppressing surface diffusion of the Fe adatoms. If
the kicked-out Ga atom subsequently diffuses away to a more stable
adsorption site, this trapping effect will be further enhanced.  Such
effects, if not kinetically hindered, will bias the growth toward
nucleating many small Fe islands.

The issue of kinetic barriers to forming Fe-Ga heterodimers can be
analyzed by considering the possible routes leading to point (d) along
the PES.  For example, the reaction pathway from point (a) to (d)
proceeds via point (c), which corresponds to the configuration shown
in Fig.~\ref{xmol3d.reactioncoords}(c). This is the transition state
(the highest energy configuration along the minimum-energy pathway)
for the reaction taking (a) into (d), and is only 0.25 eV higher than
the valley floor near (a). Part of this barrier arises from breaking
the original Ga-Ga dimer bond, but apparently most of this energy cost
is recovered by forming a more stable Fe-Ga heterodimer bond.  Another
1.0 eV is recovered at point (d) by forming, in addition, a new Fe-As
backbond to the substrate. The kicked-out Ga atom may remain bonded to
the Fe atom, but it is energetically much more favorable to break this
Fe-Ga bond (at a cost of 0.6 eV) and diffuse to a more favorable
binding site---for a net gain up to 2.0 eV per Ga atom for the limiting
case of incorporation into bulk Ga.  Comparing these energy changes upon
bond breaking, we conclude that Fe-As bonds are considerably more stable
than Fe-Ga bonds.

\begin{figure}[tb]
\centerline{\epsfig{figure=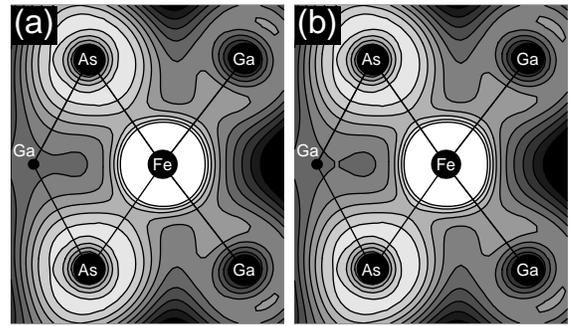,width=7.5cm}}
\caption{\label{den.pgx0y20a}
Comparison of the (a) spin-unpolarized and (b) spin-polarized valence
electron density near the Fe atom for the configuration shown in
Fig.~\protect\ref{xmol3d.reactioncoords}(b).  The contours are
logarithmically spaced; lighter contours represent higher density, and
are truncated near the Fe atom.  Projected atomic positions
are marked by black circles; their size indicates proximity to the
plotting plane. }
\end{figure}

The final adsorbate site we discuss is point (b), halfway between the
pedestal site and the dimer bridge site. At this point, the Ga-Ga
dimer bond has already broken to allow for more favorable Fe-Ga bonds
to begin forming. Without spin polarization, the energy of this
configuration is a local maximum, and represents a strong barrier to
breaking Ga-Ga dimer bonds head-on. A similar barrier is found for a
Fe adatom artificially constrained to approach the Ga-Ga dimer from
directly above, as shown in in Fig.~\ref{et-vs-z}.  For this
adsorption route from above, an energy cost of $\sim$0.4 eV must be
paid before the Ga-Ga dimer bond finally breaks; once paid, stable Fe
adsorption can occur at the bridge site.

We turn now to the differences in the PES that arise from including
spin polarization. We have repeated the calculation of the entire PES
using spin-polarized GGA, allowing for any changes in the relaxed
geometries. In general, we find only negligible changes to the relaxed
atomic positions, and so the geometries shown in
Fig.~\ref{xmol3d.reactioncoords} continue to correspond to points on
the new PES, shown in Fig.~\ref{pes.fe1.lsda}.  Comparing the
spin-unpolarized and spin-polarized energy landscapes (Figs. 1 and 4),
several features deserve comment.  The global minimum is the same for
both, corresponding the Fe-Ga heterodimer formation via Ga
``kick-out.'' With spin polarization, the pedestal site is again
unstable against adsorbate motion in any direction.

Differences are also apparent. The spin-polarized energy surface
appears more corrugated; this is due almost entirely to a strong
reduction of energies near point (b) (a local maximum without spin
polarization) and the nearby dimer bridge site (a saddle point without
spin polarization, but part of a low-energy trench with it).  Thus by
including spin-polarization, the barrier to head-on Ga-Ga dimer
breaking is reduced to zero. The same difference is observed in dimer
breaking by a Fe atom artificially constrained above the dimer: the
barrier in Fig.~\ref{et-vs-z} is eliminated by including spin
polarization---demonstrating that this effect is not limited to surface
diffusion. 

\begin{figure}[tb]
\centerline{\epsfig{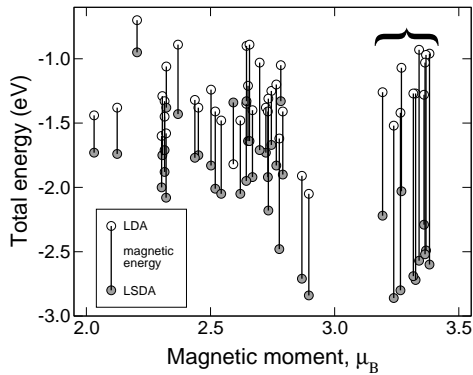}}
\caption{\label{et-vs-mom}
Relative total energies for a number of configurations used
to define the potential-energy surfaces in Figs. 1 and 4. 
The configurations within the bracket are all located near point
(b) in Figs. 1 and 4 (see text for discussion).}
\end{figure}

To understand why allowing for spin-polarization changes some parts of
the PES but not others, we consider two plausible explanations:
chemical effects and magnetic effects. By the former we mean
contributions to the total energy that depend primarily on the total
valence electron density; by the latter we mean contributions related
to the electron spin density. We focus on the configuration of
Fig.~\ref{xmol3d.reactioncoords}(b), for which the change in the PES
is particularly dramatic. In Fig.~\ref{den.pgx0y20a} we compare the
spin-unpolarized and spin-polarized total valence-electron density in
a plane containing the Fe adsorbate and its As and Ga neighbors.  In
both plots, Fe-As and Fe-Ga bonds are clearly visible, and the loss of
both the Ga-Ga dimer bond and the Ga-As backbond is obvious.  Most
importantly, there are no large differences between the spin-polarized and
unpolarized electron densities in this plane.  We conclude that the changes
in the PES due to spin-polarization cannot be attributed to changes in
the valence electron density related to chemical bonds.

\begin{figure}[tb]
\centerline{\epsfig{figure=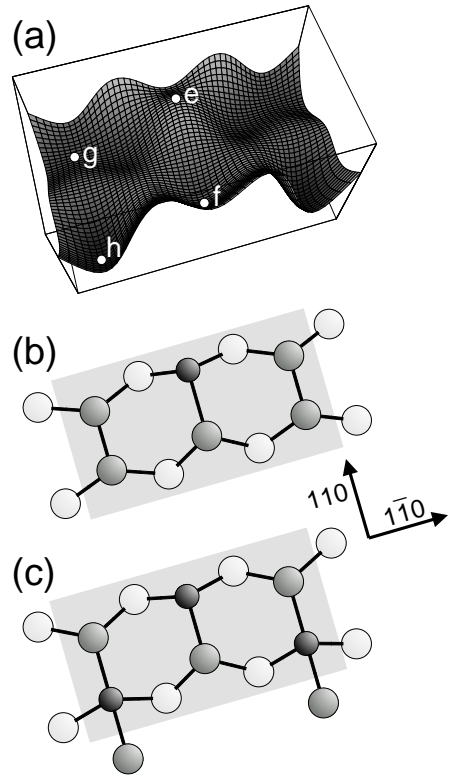,width=6.5cm}}
\caption{\label{pes+xmol.fe2}
(a) Potential energy surface for Fe on Fe/GaAs(001), calculated
with spin polarization. (b) The initial preadsorbed surface structure,
consisting of Fe-Ga heterodimers alternating
with Ga dimers. (c) Lowest-energy configuration for 1/2 monolayer Fe
coverage, corresponding to the second Fe adsorbate at point (h) in the
potential-energy surface. The shaded area marks the region 
plotted in (a). }
\end{figure}

To evaluate the role of magnetic effects in determining the shape of
the PES, we first consider how the total magnetic moment per cell
varies with the Fe adsorbate position. In Fig.~\ref{et-vs-mom} we show
the magnetic energy (the difference between spin-polarized and
unpolarized energies), as a function of total magnetic moment, for
about 50 different adsorption sites on the PES. Most of the sites have
magnetic moments between 2 and 3 $\mu_B$, giving rise to magnetic
energies between 0.4 and 0.8 eV. About ten sites have considerable
higher moments, between 3.2 and 3.4 $\mu_B$, and correspondingly
higher magnetic energies, between 1.0 and 1.6 eV. Thus these sites,
all located near point (b) on the PES, give rise to large differences
in the spin-polarized and unpolarized energy landscapes. These
differences can be attributed to the development of unusually large
magnetic moments for adsorption sites in the vicinity of energy
barriers in the spin-unpolarized PES; these sites generally involve
highly strained or partially broken bonds.  The same trend is observed for
the constrained adsorption from above (Fig.\ref{et-vs-z}). The
magnetic energy gain at the configuration corresponding to the barrier
is 2.5 eV, nearly twice as large as at the equilibrium adsorbate
height. This difference again arises from the difference in magnetic
moments: 3.9 $\mu_B$ at the barrier configuration vs.\ 3.2 $\mu_B$ at
equilibrium. Thus we conclude that magnetic effects play an important
role in determining those parts of the PES corresponding to
highly strained or partially broken bonds.

We end this section by turning briefly to another process important during
growth: the adsorption of Fe in the presence of preadsorbed Fe. We
consider 1/4 monolayer of Fe in its lowest-energy configuration, i.e.\
incorporated as Fe-Ga heterodimers with (2$\times$2) periodicity. To
simplify the discussion, we assume a starting surface from which the
kicked-out Ga atoms have detached and diffused away.  We then
recalculate the (spin-polarized) potential-energy surface for this preadsorbed
surface, again fully accounting for geometrical relaxation.

A portion of this PES is shown in Fig.~\ref{pes+xmol.fe2}(a), along
with the initial preadsorbed surface structure in
Fig.~\ref{pes+xmol.fe2}(b). The area shown is roughly twice that of
Fig.~\ref{pes.fe1.lsda}, and a general comparison of the corresponding
portions reveals the following. First, the overall corrugation of the
preadsorbed PES is smaller than for the clean surface, i.e.\
preadsorbed Fe lowers the barriers to surface diffusion.  The
locations on this PES labeled (e) through (h) correspond to global
minima on the PES of the clean surface. Point (e), located at the
position of the preadsorbed Fe, is here a local energy maximum. This
suggests that Fe-Fe bonding is not yet favorable in this low coverage
regime, at least relative to the further formation of Fe-Ga
heterodimers. Points (f) through (h) are local minima, indicating that
formation of Fe-Ga heterodimers remains favorable for the preadsorbed
surface. The global minimum, point (h), corresponds to the staggered
arrangement of Fe-Ga heterodimers shown in
Fig.~\ref{pes+xmol.fe2}(c). Occupying all of these adsorption sites
results in an Fe coverage of 1/2 monolayer.

Although we do not explicitly calculate energy surfaces for higher
coverages, we can make plausible inferences for their basic features
based on the PES of Fig.~\ref{pes+xmol.fe2}. Even for a surface
completely terminated by Fe-Ga heterodimers, the strong preference for
Fe to form backbonds to As---as evidenced by the local minimum labeled
(f) in Fig.~\ref{pes+xmol.fe2}(a)---should persist. Thus we speculate
that continued deposition of Fe will ultimately lead to Fe-Fe dimers
atop the As-terminated substrate. An obvious consequence of such a
configuration would be the release of Ga (a full monolayer in this
surface model).  Moreover, in Section III we will show that for Fe films of
several monolayers, a floating Ga (or As) adlayer can act as a
surfactant, lowering the surface energy by as much as 2 eV/(1$\times$1) cell.

\begin{figure}[btp]
\centerline{\epsfig{figure=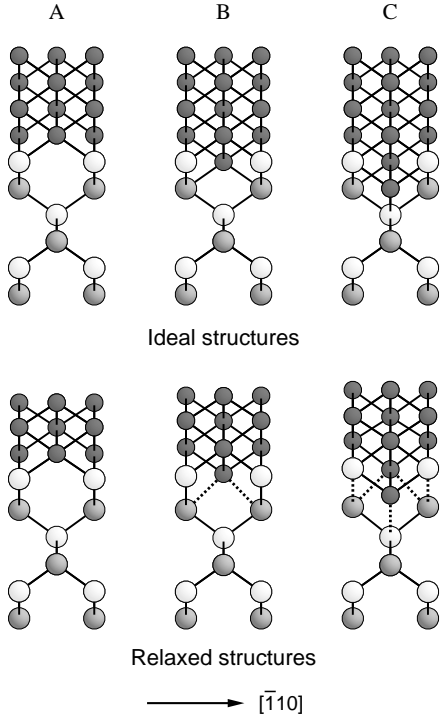,width=6.1cm}}
\caption{\label{interfacemodels}
Interface structures for the As-terminated Fe/GaAs(001) interface:
Ideal (upper panel) and relaxed structures (lower panel).  The gray
and light gray spheres represent Ga and As atoms, and the dark gray
spheres represent Fe atoms. Highly strained bonds (15--20\% longer
than ideal) are shown as dotted lines.  }
\end{figure}

To conclude this section, we have shown that the initial stages of
adsorption of Fe are dominated by strong local Fe-As chemistry. For a
surface terminated by Ga, this chemistry leads to facile breaking of
surface Ga dimers by sequential kicking-out of the two Ga atoms, first
forming Fe-Ga heterodimers and, finally, surface Fe dimers bonded to
subsurface As. Excess Ga may be released as a result of this rebonding
mechanism.  Magnetism plays an auxiliary role in the process by
lowering the potential-energy barriers to breaking apart surface Ga
dimers.  Since the formation of Fe-As bonds leads to efficient
trapping of Fe---especially if the excess Ga diffuses away---surface
diffusion of Fe will probably be strongly suppressed.

\section{Ground-state interface structure}\label{Ground_state_interface_structure}

As a complementary approach to the study of single Fe adsorption and
diffusion, we have studied epitaxial interfaces of Fe/GaAs.  In
this section, our focus is threefold: (1) to determine the stability
and magnetic character of Fe films for different interface structures
and film thickness; (2) to investigate the role of As or Ga adlayers
on Fe films, in particular the extent to which they may account for
the observed out-diffusion of substrate atoms to the surface; (3) to
examine the possibility of antiferromagnetic order as the origin of
the observed magnetic quenching of Fe films with thicknesses of just a
few monolayers.

The Fe/GaAs interface structures considered in this section are shown in
Fig.~\ref{interfacemodels} for the the As-terminated GaAs(001)
interface.  Model A is an atomically abrupt interface of bcc Fe and
zinc-blende GaAs. Because the lattice constant of bcc Fe ($a =
2.866$~\AA) is almost half that of the substrate ($a = 5.654$~\AA),
the epitaxial relationship is Fe(001)$\langle100\rangle$ $||$
GaAs(001)$\langle100\rangle$ and the atomic density of each Fe layer
is twice that of the substrate layer, with a strain of only 1.3\%.
With respect to the bcc Fe lattice, there are vacancy sites in the
adjacent GaAs lattice.  Models B and C are built by filling these
vacancy sites with Fe atoms one by one; we will refer to these
interfaces as ``partially intermixed'' and ``fully intermixed,''
respectively. For each of these models, we considered
film thicknesses ranging from 0.5 ML [one Fe atom per $(1\times1)$ of
GaAs(001)] up to 3.5 ML.   We also considered three analogous models for the
Ga-terminated interface. Both As- and Ga-terminated interface models
may have relevance for experiments with the more commonly used Ga-rich
surfaces: for example,  we have already seen in the previous section that
substitutional displacement of Ga on a Ga-rich surface leads to
an As-terminated Fe/GaAs interface. 

For the calculations in this section we again used the generalized
gradient approximation (GGA), here with an ultrasoft
pseudopotential\cite{vanderbilt90} for Fe. The electronic wave
functions and densities were described by a planewave basis with
cutoff energies of 16 Ry and 160 Ry, respectively. For each structural
model, six or seven atomic layers were used for the substrate; the
bottom As or Ga layer was passivated by pseudohydrogen atoms and fixed
during the structural relaxation.  The relative stability of each slab
is given by its formation energy,
\begin{equation}
E_{\rm form}^{\rm slab}= E_t -\sum_i N_i \mu_i,
\end{equation}
where $E_t$ is the total energy of the slab, $N_i$ the number of atoms of each
chemical type, and $\mu_i$ their chemical potentials.
To eliminate the contribution of the pseudohydrogen layer to the formation
energy, we define the formation energy of each structural model with respect to a
common reference structure,
\begin{equation}
E_{\rm form}^{\rm model}= E_{\rm form}^{\rm slab}-E_{\rm form}^{\rm ref}/2.
\end{equation}
The reference structure is an ideal GaAs slab passivated on both sides
(hence the factor 1/2) by pseudohydrogen atoms.
Assuming the thermodynamic equilibrium condition,
$\mu_{\rm GaAs}=\mu_{\rm As}+\mu_{\rm Ga}$, and taking 
$\mu_{\rm GaAs}$ and $\mu_{\rm Fe}$ from bulk structures,
the formation energy is expressed as a function of $\mu_{\rm As}$ only
within the thermodynamically allowed range 
$\mu_{\rm GaAs}-\mu_{\rm Ga(bulk)}<\mu_{\rm As}<\mu_{\rm As(bulk)}$.
Here, the lower (upper) limit corresponds to the Ga-rich (As-rich)
environment, and $\mu_{\rm Ga(bulk)}$ and $\mu_{\rm As(bulk)}$ are 
again determined from their bulk structures. 

\begin{figure}[tb]
\centerline{\epsfig{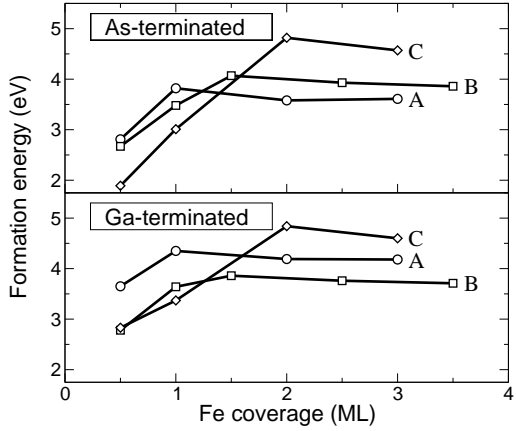}}
\caption{\label{eformation}
Formation energies [in eV/(1$\times$1)-cell]
of Fe films on the GaAs(001) surface given as a function of interface
structure (A,B,C) and Fe coverage.  The formation energies given are taken at the
center of the thermodynamically allowed range of the As chemical
potential: in the As-rich limit condition the formation energies
decrease (increase) by 0.16 eV for the As-terminated
(Ga-terminated) Fe/GaAs interface. }
\end{figure}

Figure~\ref{eformation} shows the calculated formation energies as a
function of Fe film thickness.  We first discuss Fe films with the
As-terminated interface.  At a Fe coverage of 0.5 ML model C is
most stable, while models A and B are $\sim$1 eV higher.  This
energetic ordering persists up to 1 ML coverage, but changes above
that point: the formation energy of model C increases by 1.8 eV
from 1 ML to 2 ML, while that of the A and B become nearly
independent of coverage.  After the formation of 2 ML, the formation
energy of all three models does not change much by adding more Fe
layers.  Thus, the low-energy interface structure turns out to be
model A---the abrupt interface.  For Fe films with the
Ga-terminated interface, a similar trend applies but with a different
outcome.  At 0.5 ML, models B and C are equally stable, and at 1
ML model C is in fact the most stable.  With additional Fe layers,
model C becomes very unstable, similar to the case of the
As-terminated interface. At higher coverages, however, model B---the
partially mixed interface---is most stable.

\begin{figure}[tb]
\centerline{\epsfig{figure=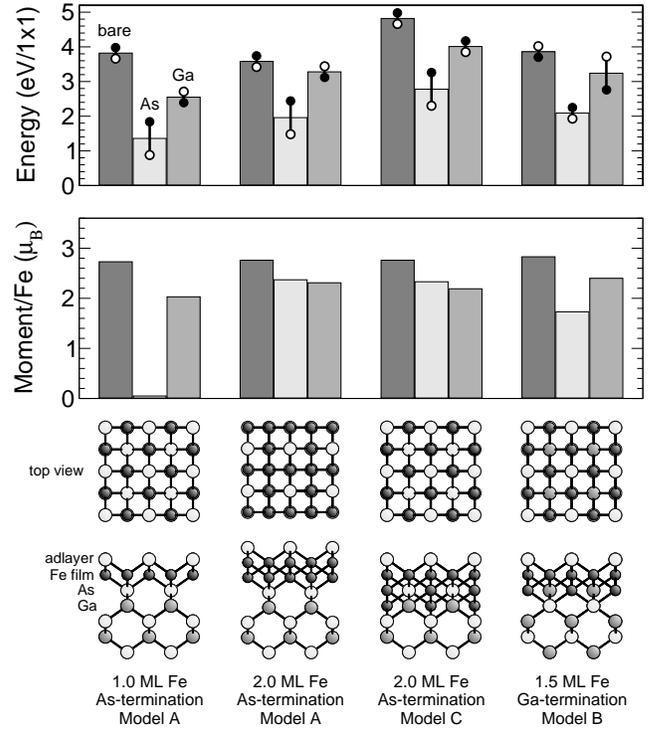,width=8.5cm}}
\caption{\label{adlayerenergies}
Formation energies and magnetic moments for four selected
adlayer structures.  The Fe film thickness, substrate termination, and
interface model is listed underneath each structure. Formation
energies and magnetic moments are shown for bare Fe films 
and for As- and Ga-adlayers.  The formation energies are calculated
at the center of the thermodynamically allowed range of the As
chemical potential ($\mu_{\rm As}$).  The changes at the limiting
values of $\mu_{\rm As}$ are indicated by small circles (filled circles
correspond to the Ga-rich limit, open circles to the As-rich limit). }
\end{figure}

These results can be understood as a competition between maximizing
the coordination of Fe atoms and minimizing the concentration of atoms
in the interfacial region.  At low Fe coverage ($\le 1$ ML), 
model C is energetically most favored, because Fe atoms can maximize
their coordination with substrate atoms.  At higher coverages, where
the interface is well defined, this arrangement becomes unstable
relative to the less intermixed interfaces. The reason for this
crossover is the two extra Fe atoms per ($1\times1$) surface cell
(relative to model A).  The excess electrons from the extra Fe
atoms fill antibonding orbitals and thus weaken the interface bonding.
This weakened bonding is also evident in the relaxed interface
separations: for both As- and Ga-terminated interfaces, the position of
the first Fe monolayer is $\sim$0.8~{\AA} higher in model C and
$\sim$0.4~{\AA} higher in model B, compared to model A (see
the relaxed structures in Fig.~\ref{interfacemodels}).  Similar
reasoning explains why model B is more stable than model A for
the Ga-terminated interface, while the opposite holds for the
As-terminated interface.

The present results emphasize that the low-energy atomic structure at
low Fe coverage may not be extrapolated to high coverages.  This
should be true also in real growth situations, since Fe atoms at low
coverages would take positions one or two layers deep so as to
maintain maximal coordination.  Further deposition of Fe leads to an
partially mixed interface that ultimately becomes unstable. Therefore,
it is likely that substantial rearrangement of the atomic structure
occurs during the film growth, provided the temperature is
sufficiently high.  This may partly account for the
experimental observation
\cite{kneedler97b,chamber86,monchesky99} that substrate atoms, 
especially As atoms, diffuse out to the surface during the Fe growth
even for room-temperature deposition.  

To follow this reasoning, we examined the effect of a Ga or As adlayer
on top of the Fe film.  In Fig.~\ref{adlayerenergies} we compare the
formation energies of four selected interface structures with and
without an adlayer.  Both Ga and As adlayers stabilize the surface
substantially, regardless of the interface structure and film
thickness. This is not surprising, since adlayers increase the
coordination of surface Fe atoms. The energy gain from As adlayers is
particularly large, amounting to more than 1.5 eV/$(1\times1)$ in most
cases.  Adlayers on 1 ML of Fe show strong dependence on the adatom
site: the low-energy site is the position extended from the substrate
GaAs lattice, suggesting a strong covalent bonding between the
substrate and the adlayer, mediated by the intervening Fe layer.

\begin{figure}[tb]
\centerline{\epsfig{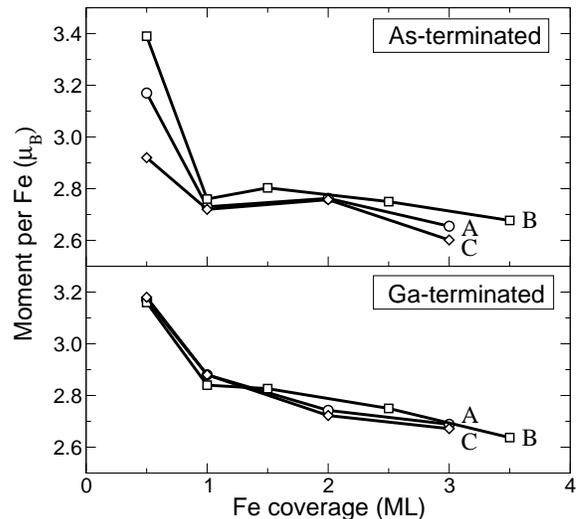}}
\caption{\label{moment}
Magnetic moments of Fe films on GaAs(001) as a function of interface
type and Fe coverage. }
\end{figure}
 
The results for adlayer structures, together with the interface
energetics, provide strong theoretical evidence that substitutional
adsorption and/or atomic exchange is an essential process during the
Fe growth---a finding which was anticipated by
experiments\cite{chamber86} and which we explicitly demonstrated 
in Section \ref{Initial_stages_of_growth}.  By substitutional
adsorption processes, Fe atoms can maximize their coordination at
every stage of the growth. These processes result in the segregation
of substrate atoms to the surface, and simultaneously facilitate
optimal interfacial atomic densities and thus relatively stable
interface structures.  Our calculations have shown that such processes
are inevitable and that the energy gain from them is quite
substantial.

\begin{table}
\caption{\label{tab_spin}
Atom-resolved spin moments (in $\mu_B$) for Fe/GaAs interface
structures.  The layer number is given with respect to the top
substrate layer.  The induced spin moments of substrate atoms are
given with a species label.  }
\begin{tabular}{cccc}
Layer &  As:A 3.0 ML & Ga:B 3.5 ML       & As:A 2.0 ML $\backslash$ As \\
\hline
  3   &  3.03, 3.00  & 3.02, 2.99        & $-$0.08 (As)                \\
  2   &  2.51, 2.38  & 2.49, 2.43        & 2.52, 2.42                  \\
  1   &  2.54, 2.49  & 2.49, 2.42        & 2.40, 2.29                  \\
  0   & $-0.04$ (As) & 2.72, $-0.06$ (Ga)& $-$0.04 (As)                \\
 $-1$ &  0.02 (Ga)   & 0.00 (As)         &  0.02 (Ga)                  \\
 $-2$ &  0.01 (As)   & 0.03 (Ga)         &  0.02 (As)                  \\
 $-3$ &  0.00 (Ga)   & 0.01 (As)         &  0.00 (Ga)                  \\
\end{tabular}
\end{table}

We now turn to magnetic properties of the interface structures.
Calculated spin moments (per Fe atom) of various model structures are given in
Figs.~\ref{adlayerenergies} and \ref{moment}.  For bare Fe films
without Ga or As adlayers, average Fe spin moments are much enhanced
compared to the calculated bulk value for Fe of 2.33 $\mu_B$.
Fig.~\ref{moment} shows that for films of 3 ML this enhancement
is still substantial, and relatively insensitive to the specific interface
structure or substrate termination. For Fe films with adlayers,
Fig.~\ref{adlayerenergies} shows that both As- and Ga-adlayers
suppress total magnetic moments, by as much as 1 $\mu_B$ in several
cases.

In order to examine local variations in the total magnetic moments, we
calculated partial moments within a sphere centered on each atom, in
analogy to the muffin-tin sphere in all-electron approaches. We used a
sphere radius of 1.2 {\AA}, which is slightly shorter than the minimum
bond length between atoms.  Table~\ref{tab_spin} shows the results for
three selected structures.  Local Fe moments are significantly
enhanced ($\sim$3.0 $\mu_B$) at the surface layer. For the relatively
thin Fe films considered in this work, the buried and interfacial Fe
layers also have sizable enhancement.  The adlayer suppresses spin
moments not only of the top-layer Fe atoms but also slightly of the
second-layer Fe atoms.  Both the enhancement and suppression can be
understood in terms of changes in the Fe $d$-band width due to
changes in coordination and symmetry.  On the other hand, substrate
atoms have small induced moments up to 3--4 layers deep, always with
negative spin moments for the interfacial atoms.  The sign of spin
moments changes to positive at deeper layers, suggesting the formation
of a spin-density wave.

\begin{figure}[tb]
\centerline{\epsfig{figure=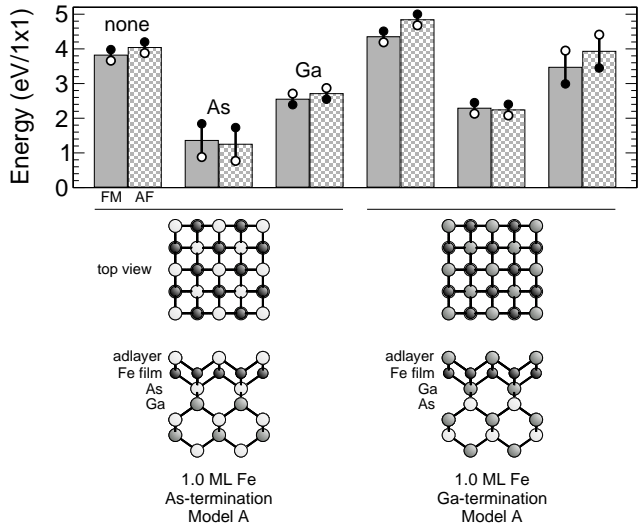,width=8.5cm}}
\caption{\label{eform+mom+xmol}
Formation energies of ferromagnetic (FM) and antiferromagnetic (AF) Fe
films in two different Fe/GaAs structures (As- and Ga-terminated abrupt
interfaces). Three types of adlayers are considered (none, As, Ga).}
\end{figure}
 
Recent magnetic measurements have observed a delayed onset of the
magnetic phase at $\sim$2 ML for Fe films grown on the
Ga-rich GaAs(001)-$(2\times6)$ and $(4\times2)$ surfaces. \cite{bensch01} Of
the various surface structures we have considered here, one shows a
similar quenching effect. The As-terminated interface with 1 ML of Fe
(the first structure in Fig.~\ref{adlayerenergies}) has a zero net
moment when capped by an As adlayer. Since this structure also has the
lowest formation energy for an Fe coverage of 1 ML, it might be
anticipated that the delayed onset is due to the quenching of
magnetism from the strong covalent bonding between the As atoms via Fe
atoms.  However, there is another possibility, namely the formation of
antiferromagnetic (AF) order.  We have considered the possibility of
AF order for the simplest structures with 1 ML of Fe in the same
plane, so that one Fe atom has spin up and the other down. The
results, summarized in Fig.~\ref{eform+mom+xmol}, indicate that films
with an As adlayer become more stable by 0.05--0.09 eV/$(1\times1)$
upon the formation of AF order, while bare films and films with a Ga
adlayer become unstable by 0.2--0.5 eV/$(1\times1)$.  We note that the
lowest energy structure with quenched magnetism is unstable toward AF
order.  We propose on this basis that the observed delayed magnetic
onset is due to the initial formation of AF order, and tightly
correlated with the out-diffusion of As atoms to the surface.

While the microscopic origins driving the formation of AF order are
not clear in detail, it appears that the surface As atoms mediate the
AF order between Fe atoms.  A related observation is that the
AF-stabilized surface structure is locally similar to the tetragonal
Fe$_2$As structure in the AF ground state: the unit cell consists of
bimolecular units with two Fe atoms in the same plane, plus one As and
one Fe atom on each side of the Fe plane. \cite{wyckoff}  Thus it is
plausible that AF order in Fe/GaAs interfaces at low coverage arises
from the formation of the Fe$_2$As-like structures.  This local
similarity is broken by additional Fe adsorption, so that 
ferromagnetic order is ultimately favored.

To conclude this section, we have shown that the low-energy
interface structure is different for low and high Fe coverages.  The
addition of Ga- or, especially, As-adlayers substantially lowers the
surface formation, providing a theoretical basis for the
experimentally observed out-diffusion of substrate atoms to the
surface.  We have shown that Fe films on the GaAs substrate usually
assume a ferromagnetic ground state even at very low coverages, while
for certain cases an As adlayer can induce antiferromagnetic order in
the Fe film.  Since the latter cases generally have lower
formation energies and are thus likely to form, we propose that the
observed magnetic quenching for very low Fe coverage may be due
to the formation of antiferromagnetic order.
 
\section{Summary and future directions}\label{Summary}

We have taken two complementary approaches to understanding the
nucleation and growth of Fe on GaAs---one focusing on the behavior of
single Fe adatoms deposited on clean and partially preadsorbed
GaAs(001), the other focusing on the interface structure of complete
Fe films at coverages up to several monolayers. Although a detailed
growth history of Fe/GaAs interfaces cannot yet be described, our
studies suggest four generic features that may play an important role.

First, for low coverages we have identified a very strong driving
force for Fe to be highly coordinated.  For single adatoms, this
tendency is strong enough to spontaneously break surface Ga-Ga and
Ga-As bonds in order to form Fe-As bonds.  Although magnetic effects
do not dominate this chemistry, they play an interesting auxiliary
role by essentially removing energetic barriers from the reaction
pathways.  For half-monolayer films constrained by symmetry, these
local surface chemical reactions are not possible, and the Fe atoms
respond by occupying subsurface sites with high coordination to either
Ga or As.

A second generic feature is the crossover from a preference for
strongly intermixed films to less intermixed or even abrupt
films. This crossover occurs between one and two ML for both As- and
Ga-terminated interfaces.  Its origin is not magnetic: the films exhibit
moments per atom larger than the bulk value even for very low
coverages, and the moments converge to their bulk value long after the
crossover occurs.  Rather, it arises from the competition between
maximizing the coordination of Fe atoms (which favors intermixing) and
minimizing the amount of excess interfacial Fe (which favors abrupt
interfaces).  For well-defined Fe films of 2 ML or more, the latter
effect dominates and sharper interfaces become energetically
preferred.

A third finding, common to all interfaces we have studied, is that Ga-
and As-adlayers dramatically reduce the formation energies of Fe
films. This stabilizing effect occurs for both Ga- and As-terminated
interfaces, for both intermixed and abrupt interfaces, and for all
film thicknesses considered.  It is especially striking for
As-adlayers, which can reduce the film formation by as much as 50\%.
We also find that this stabilization is generally accompanied by a
suppression of the total magnetic moment of the film; since this is
due to reduced local moments in the topmost layer or two, the effect
is largest for thinner films.  Indeed, for 1-ML Fe films with an
As-adlayer, antiferromagnetic order can be more stable than
ferromagnetic order.

Taken together, these three generic features collectively imply a
fourth: the diffusion of Ga or As atoms from the interface to the
surface of the Fe film.  We have shown explicitly how Ga can be
released by adsorption of individual Fe adatoms and the subsequent
``kick-out'' of Ga from surface dimers. Even for GaAs surface
reconstructions that do not consist of Ga dimers,\cite{lee00} we
speculate that the same strong Fe-As chemistry would again lead to the
release of surface Ga atoms.  We have also shown that during the
growth---between one and two ML---a spontaneous rearrangement of the
interface morphology is likely to occur, again leading to the release
of either Ga or As. Although we do not speculate about the details of
this atomic rearrangement, we have also shown that ultimately the
liberated Ga or As is likely to play the role of a floating surfactant
layer.

Finally, we mention a possible avenue for further research. One
difficulty with theoretical studies of interface is the paucity of
macroscopic observables that can be directly related to the
microscopic interface structure. Schottky barriers are extremely
sensitive probes of interface structure, varying by as much 25\% for
local changes in interface geometry.\cite{heslinga90} Schottky barrier
heights have been measured in Fe/GaAs interfaces to be of order 0.7
eV,\cite{bland01} and thus represent a useful probe of interface
microstructure.  Moreover, Schottky barriers may actually be a
necessary ingredient for circumventing the intrinsic limitations on
spin injection from a ferromagnetic metal into a semiconductor. A
theoretical understanding of their dependence on Fe/GaAs interface
structure---including substrate termination and reconstruction, degree
of intermixing, and magnetic character---would be a great asset.

\section{Acknowledgments}

This work was funded in part by ONR and in part by the 
Deutsche Forschungsgemeinschaft. Computational work was supported in
part by a grant of HPC time from the DoD Major Shared Resource Center
ASCWP. S.C.E. and S.H.L. acknowledge generous support from the
Alexander von Humboldt Foundation.


\begin{references}
\bibitem[*]{permanentaddress} Permanent address: 
Computational Science and Engineering Center,
Samsung Advanced Institute of Technology,
P.O. Box 111, Suwon 440-600, South Korea
\bibitem{prinz98} G. A. Prinz, Science {\bf 282}, 1660 (1998)
\bibitem{fiederling99} R. Fiederling, M. Keim, G. Reuscher, W. Ossau, G. Schmidt, A. Waag, and L. W. Molenkamp, Nature (London) {\bf 402}, 787 (1999).
\bibitem{jonker00} B. T. Jonker, Y. D. Park, B. R. Bennett, H. D. Cheong, G. Kioseoglou, and A. Petrou, Phys. Rev. B {\bf 62}, 8180 (2000). 
\bibitem{ohno99} Y. Ohno, D. K. Young, B. Beschoten, F. Matsukura, H. Ohno, and D. D. Awschalom, Nature (London) {\bf 402}, 790 (1999).
\bibitem{haas70} C. Haas, IBM J. Res. Dev. {\bf 14}, 282 (1970).
\bibitem{prinz81} G. A. Prinz and J. J. Krebs, Appl. Phys. Lett. {\bf 39}, 397 (1981).
\bibitem{hammar99} P. R. Hammar, B. R. Bennett, M. J. Yang, and M. Johnson, Phys. Rev. Lett {\bf 83}, 203 (1999).
\bibitem{xu99} Y. B. Xu, D. J. Freeland, E. T. M. Kernohan, W. Y. Lee, M. Tselepi, C. M. Guertler, C. A. F. Vaz, J. A. C. Bland, S. N. Holmes, N. K. Patel, and D. A. Ritchie, J. Appl. Phys. {\bf 85}, 5369 (1999).
\bibitem{krebs87} J. J. Krebs, B. T. Jonker, and G. A. Prinz, J. Appl. Phys. {\bf 61}, 2596 (1987).
\bibitem{filipe97} A. Filipe, A. Schuhl, and P. Galtier, Appl. Phys. Lett. {\bf 70}, 129 (1997).
\bibitem{anderson95} G. W. Anderson, M. C. Hanf, and P. R. Norton, Phys. Rev. Lett {\bf 74}, 2764 (1995).
\bibitem{zoelfl97} M. Z\"{o}lfl, M. Brockmann, M. Kohler, S. Kreuzer, T. Schweinbock, S. Miethaner, F. Bensch, and G. Bayreutherp, J. Mag. Magn. Mat. {\bf 175}, 16 (1997).
\bibitem{kneedler97a} E. M. Kneedler and B. T. Jonker, J. Appl. Phys. {\bf 81}, 4463 (1997).
\bibitem{xu98} Y. B. Xu, E. T. M. Kernohan, D. J. Freeland, A. Ercole, M. Tselepi, and J. A. C. Bland, Phys. Rev. B {\bf 58}, 890 (1998).
\bibitem{schmidt00} G. Schmidt, D. Ferrand, L. W. Molenkamp, A.T. Filip, and B. J. van Wees, Phys. Rev. B {\bf 62}, R4790 (2000).
\bibitem{tang00} H. X. Tang, F. G. Monzon, R. Lifshitz, M. C. Cross, and M. L. Roukes, Phys. Rev. B {\bf 61}, 4437 (2000).
\bibitem{rashba00} E.I. Rashba, Phys. Rev. B {\bf 62}, R16267 (2000).
\bibitem{hirohata00} A. Hirohata, Y. B. Xu, C. M. Guertler, and J. A. C. Bland, J. Appl. Phys. {\bf 87}, 4670 (2000).
\bibitem{zhu01} H. J. Zhu, M. Ramsteiner, H. Kostial, M. Wassermeier, H.-P. Sch\"{o}nherr, and K. H. Ploog, Phys. Rev. Lett {\bf 87}, 016601 (2001).
\bibitem{hanbicki02} A. T. Hanbicki, B. T. Jonker, G. Itskos, G. Kioseoglou, and A. Petro, Appl. Phys. Lett. {\bf 80}, 1240 (2002).
\bibitem{kratzer01} P. Kratzer and M. Scheffler, Computing in Science and Engineering {\bf 3}, 16 (2001).
\bibitem{kneedler97b} E. M. Kneedler, B. T. Jonker, P. M. Thibado, R.J. Wagner, B. V. Shanabrook, and L. J. Whitman, Phys. Rev. B {\bf 56}, 8163 (1997).
\bibitem{lee00} S. H. Lee, W. Moritz, and M. Scheffler, Phys. Rev. Lett {\bf 85}, 3890 (2000).
\bibitem{perdew92} J. P. Perdew, J. A. Chevary, S. H. Vosko, K. A. Jackson, M. R. Pederson, D. J. Singh, and C. Fiolhais, Phys. Rev. B {\bf 46}, 6671 (1992).
\bibitem{bockstedte97} M. Bockstedte, A. Kley, and M. Scheffler, Computer Physics Commun. {\bf 107}, 187 (1997).
\bibitem{vanderbilt90} D. Vanderbilt, Phys. Rev. B {\bf 41}, 7892 (1990).
\bibitem{chamber86} S. A. Chambers, F. Xu, H. W. Chen, I. M. Vitomirov, S. B. Anderson, and J. H. Weaver, Phys. Rev. B {\bf 34}, 6605 (1986).
\bibitem{monchesky99} T. L. Monchesky, B. Heinrich, R. Urban, K. Myrtle, M. Klaua, and J. Kirschner, Phys. Rev. B {\bf 60}, 10 242 (1999).
\bibitem{bensch01} F. Bensch, G. Garreau, R. Moosbuhler, G. Bayreuther, and E. Beaurepaire, J. Appl. Phys. {\bf 89}, 7133 (2001).
\bibitem{wyckoff} R. W. G. Wyckoff, {\it Crystal Structures}, 2nd Ed., Vol. 1, (Interscience Publishers, New York, 1963), p. 360.
\bibitem{heslinga90} D. R. Heslinga, H. H. Weitering, D. P. van der Werf, T. M. Klapwijk, and T. Hibma, Phys. Rev. Lett {\bf 64}, 1589 (1990).
\bibitem{bland01} J. A. C. Bland, A. Hirohata, C. M. Guertler, Y. B. Xu, and M. Tselepi, J. Appl. Phys. {\bf 89}, 6740 (2001).

\end{references}
\end{document}